\documentclass[aps,amsfonts]{revtex4}
\usepackage{epsfig}
\usepackage{graphicx}
\newcommand{\s}{\hspace{-0.18cm}\slash}
\begin{document}
\preprint{USACH-FM-00-08 }
\title{CENTRAL CHARGES AND EFFECTIVE ACTION AT FINITE TEMPERATURE AND
DENSITY}
\author{J. Gamboa$^{1}$\thanks{E-mail: jgamboa@lauca.usach.cl}, J. L\'opez-Sarri\'on$^{1,2}$
\thanks{E-mail: justo@fisica.usach.cl}, M.~Loewe$^{3}$ \thanks{E-mail: mloewe@fis.puc.cl} and
F. M\'endez$^{4}$\thanks{E-mail: fernando.mendez@lngs.infn.it}}
\address{$^{1}$Departamento de F\'{\i}sica, Universidad de Santiago de Chile,
Casilla 307, Santiago 2, Chile
\\
$^{2}$Departamento de F\'{\i}sica Te\'orica, Universidad de Zaragoza, Zaragoza 50009, Spain
\\
$^{3}$Facultad de F\'{\i}sica, Pontificia Universidad Cat\'olica de Chile, Casilla 306, Santiago 22, Chile 
\\
$^{4}$INFN, Laboratori Nazionali del Gran Sasso, SS, 17bis, 67010 Asergi (L'Aquila),  Italy}
\affiliation{} 
\date{\today}

\begin{abstract}
The current algebra for gauge theories like QCD at finite temperature and density is studied. We
start considering, the massless Thirring model at finite temperature and density, finding an
explicit expression for the current algebra. The central charge only depends on the
coupling constant and there are  not new effects due to temperature and density. From this
calculation, we argue how to compute the central charge for $QCD_4$ and we argue why the
central charge in four dimensions
could be modified by finite temperature and density.
\end{abstract}
\maketitle
\narrowtext
\section{Introduction}

Anomalies appear at quantum level because some classical symmetries are not
preserved. There are some dramatic experimental consequences from the anomalies as,  for example in the 
$\pi^0 \rightarrow 2 \gamma$ decay \cite{pes}.

Although there are finite temperature and density corrections associated to the ultraviolet sector,
the most important contribution to those corrections, probably, will come from nonperturbative
effects in the infrared sector\cite{ka,lo,we,fp,da1,re}.  This last fact is particularly relevant when phenomena
such as hadronization and deconfinement are considered.

If, formally,  we consider  $Z$ as  the generating functional 
\[
Z = \int {\cal D} A_\mu ~e^{-S[A]} ~Z_F [A], 
\]
describing a general gauge field theory, where $S[A]$ is the kinetic term for the gauge fields and $Z_F [A] $ is the generating functional for the fermionic fields minimally coupled to the $A_\mu$, then   the fermionic vector current $J_\mu$ in the chiral limit is defined as 
\begin{equation}
\frac{\delta Z_F[A]}{ \delta A^\mu} =\left< J_\mu  \right>, \label{1}
\end{equation}
where $J_\mu = {\bar \psi} \gamma_\mu \psi$.

Then, the relation
\begin{eqnarray}
&&\frac{\delta^2 Z_F[A]}{\delta A^{\mu}({\bf x}) \delta A^\nu ({\bf x}^{'})} - \frac{\delta^2 Z_F[A]}{\delta A^{\nu}({
\bf x^{'}}) \delta A^\mu ({\bf x})} = \nonumber
\\
&=& \left< \left[ J_\mu ({\bf x}), J_\nu ({\bf x}^{'})\right] \right>,  \label{2}
\end{eqnarray}
provides, in principle, an explicit calculation procedure for the current algebra.

This last statement needs some clarifications. Indeed, in the past many people computed the current algebra \cite{joh} using canonical methods and point splitting regularization. However, the novelty of our approach is that, firstly, we compute the  fermionic determinant at finite  temperature and density and, later, we compute and show the universal character of the current algebra.  

The direct calculation of the current algebra in four-dimensions involves many technicalities and
the appearence of divergences. It is, therefore, not easy to recognize the temperature and density
corrections from such direct calculation. 

These considerations motived us to start considering,
initially, a more simple problem in order to get some physical insight for the four-dimensional case
which will be considered later.

Thus,  we will start  considering the massless Thirring model at $T=0$ and we compute
explicitly the central charge \cite{elcio}. From this calculation,  we will be able to extract also the current
algebra structure for finite temperature and density and we will show that the central charges for
$T=0$ ($c_{T=0}$) and $T\neq 0$ ($c_{T\neq 0}$)  satisfy
\begin{equation}
c_{T=0} = c_{T\neq 0},  \label{3}
\end{equation}
{\it i.e.} finite temperature and density do not induce corrections to the central charge. Although the temperature independence could be argued 
given physical arguments, the density independence is nontrivial as we will discuss in the conclusions. 

This letter is organized as follows: in section 2, the current algebra for the Thirring model at
$T=0$ is considered. In section 3, we discuss  temperature and density effects in the Thirring
model.  In section 4 we argue how to extend  the previous results to higher dimensions. Finally, in section 5 we give the conclusions and outlook. 
\section{ Central charge in  the Thirring model}

In order to introduce the calculation method, let us start discussing how to compute the current
algebra for the free case.

One start considering the lagrangian
\begin{equation}
{\cal L} = \bar\psi\left(i\partial\hspace{-2.4mm}\slash\hspace{1.4mm}-gA\hspace{-2.4mm}\slash
\hspace{1.4mm}\right)\psi, \label{5}
\end{equation}
where $A_\mu$ is an auxiliary field which will vanish at the end of the calculation.

Then, the  partition function in the euclidean space is given by
\begin{equation}
Z_A =  \int {\cal D} {\bar \psi} {\cal D}\psi ~e^{-S[A]}.  \label{6}
\end{equation}

From (\ref{6}),  the vector current becomes
\begin{eqnarray}
\frac{\delta Z_A}{\delta A^\mu}\bigg|_{A=0}  &=& \int {\cal D} {\bar \psi} {\cal D}\psi ~\bar\psi(x)
\gamma_\mu\psi(x)~e^{-S[A]} \nonumber
\\
&=& g \left< J_\mu (x)\right>_{A=0},  \label{7}
\end{eqnarray}
and, therefore, the Ward identity becomes
\begin{equation}
\partial_\mu \left< J_\mu (x)\right>_{A=0} = 0. \label{co}
\end{equation}

Although in this free case, this identity is satisfied identically, in the interacting case this
conservation law could fail. Indeed, the procedure discussed above implies to compute the
fermionic determinant and, therefore, a  regularization scheme must be used.

If one consider an interacting quantum field theory involving classical gauge and chiral
symmetries, then one cannot retain simultaneously both symmetries.  For example, if we regularize 
the fermionic determinant with a gauge invariant regularization scheme then, the chiral symmetry
is explicitly violated and vice versa.

Some questions, relevants in this context, are for example, which are the implications for the current algebra calculation?, will be this
algebra modified due to temperature and density effects?, which are the modification in four
dimensions? and so on.

From (\ref{7}), one could compute the current algebra by means of 
\begin{equation}
\frac{\delta^2 Z_A}{\delta A^\mu ({\bf x}) \delta A^\nu ({\bf y})} -  \frac{\delta^2 Z_A}{\delta A^\nu ({\bf
y}) \delta A^\mu({\bf x})} = \left< \left[ J_\mu ({\bf x}), J_\nu ({\bf y})\right] \right>, \label{8}
\end{equation}
using different alternative regularization procedures and not necessarily the splitting point regularization.

We will consider below the  Thirring model
\begin{equation}
{\cal L}_T= \bar\psi i\partial\hspace{-2.4mm}\slash\hspace{1.4mm}\psi -g^2(\bar\psi\gamma_\mu
\psi)(\bar\psi\gamma^\mu\psi),  \label{9}
\end{equation}
and will compute the associated current algebra.

In order to do that, it is more convenient to use the linearized form of (\ref{8}), {\it v.iz.}
\begin{equation}
{\cal L}_A=\bar\psi\left(i\partial\hspace{-2.4mm}\slash\hspace{1.4mm}-gA\hspace{-2.4mm}\slash
\hspace{1.4mm}\right)\psi+
\frac{1}{4}A^2, \label{10}
\end{equation}
which is equivalent to (\ref{9}) after integrating $A_\mu$.

This last fact implies to change $A_\mu$ into 
\begin{equation}
A_\mu(x) \rightarrow 2g\bar\psi(x)\gamma_\mu\psi(x)=2gJ_\mu(x), \label{11}
\end{equation}
and, as a consequence, (\ref{9}) and (\ref{10}) are equivalent.

In this interacting case,  the vector current in the sense of (\ref{7}) is
\begin{equation}
 \frac{\delta Z_A}{\delta A_\mu}\bigg|_{A \neq 0} = \frac{1}{2}  \left< A_\mu \right> - g \left< J_\mu\right>, \label{122}
 \end{equation}
 where the first term in RHS in (\ref{122}) is the correction to the conservation law (\ref{co})  due
 to the interaction. We would like to emphasize, however, that the notation $\left< A_\mu \right> $ is formal because the expectation value is computed integrating the fermionic fields. However, this notation will be useful in the next formula where -by using an appropriated trick- the integral in $A_\mu$ will appear explicitly. 

 Although classically the relations $\partial_\mu \left< A_\mu \right> =0$ and
 $ \partial_\mu \left< J_\mu\right>=0 $ hold, at the quantum level the situation is quite different

  Indeed, in order to compute the current algebra one use the identity

\begin{eqnarray}
&&\left< A_\mu (x) A_\nu (x^{'})\right> =  \int {\cal D} {\bar \psi} {\cal D}\psi{\cal D} A ~A_\mu (x) A_\nu (x^{'})~
e^{-S[A]} \nonumber
\\
&=&  \int {\cal D} {\bar \psi} {\cal D}\psi~e^{-S_T} \int {\cal D} A ~\left[ A_\mu (x) - 2 g J_\mu (x)
\right] \times  \nonumber
\\
&\times& ~\left[ A_\nu (x^{'}) - 2 g J_\nu (x^{'})\right]~e^{-\frac{1}{4} \int d^2x \left( A_\mu - 2 g
J_\mu\right)^2} + \nonumber
\\
&4& g^2  \int {\cal D} {\bar \psi} {\cal D}\psi~J_\mu (x) J_\nu (x^{'})~e^{-S_T}. \label{12}
\end{eqnarray}

The first term in the RHS is  just a gaussian integral as can be checked using the change of
variables
\[
A_\mu\rightarrow A_\mu^\prime=A_\mu-2g\,J_\mu, 
\]
and, therefore, one obtain
\begin{equation} \int{\cal D}A^\prime\,A^\prime_\mu(x)A_\nu^\prime(x^\prime)\,e^{-\frac{1}{4}\int d^2x\,A^{\prime 2}} \left\langle 1\right\rangle_T= 2\delta_{\mu\nu}\delta^2(x-x^\prime)\left\langle 1\right\rangle_T,
\nonumber
\end{equation}
where in the last step  $\left\langle 1\right\rangle_T$ is
a notation for
\[
\left\langle 1\right\rangle_T = \int {\cal D} {\bar \psi} {\cal D}\psi~e^{-S_T}.
\]

The second term in the RHS (\ref{12}), contains  the product of fermionic currents
evaluated at two differents points. However, as we are interested in the current commutator, the term $ \left\langle 1\right\rangle_T$ is cancelled out and, therefore, the only term that finally remains is the commutator $<[J_\mu (x),J_\nu (x^{'})]>$. Notice that we are unable to calculate directly this term. However, the LHS of (\ref{12}) can be computed through the fermionic determinant \cite{fuckiw}   
\begin{equation}
 \int {\cal D}A\,A_\mu(x)A_\nu(x^\prime)e^{-\frac{1}{4}\int  d^2x\,\left[\left(\frac{g^2}{\pi}F_{
\sigma\rho}\Box^{-1}F^{\sigma\rho}\right) +A^2\right]}. \label{14}
\end{equation}

In the momentum space (\ref{14}) becomes
\begin{eqnarray}
\int \frac{d^2k}{(2\pi)^2}&&e^{-ik(x-x^\prime)}\int {\cal D}\tilde A\,\tilde A_\mu(-k)\tilde A_\nu(k) e^{-\frac{1}{2}\int d^2k^\prime \tilde A_\sigma(-k^\prime)
G_{\sigma\rho}(k^\prime)\tilde A_\rho(k^\prime)} =\nonumber
\\
&=&\int d^2k \,G^{-1}_{\mu\nu}(k)e^{-ik(x-x^\prime)}, \label{15}
\end{eqnarray}
where the inverse of $G_{\mu\nu}(k) $ is
\begin{equation}
G_{\mu\nu}^{-1}(k)=2\left\{\frac{\frac{2g^2}{\pi}}{1-\frac{2g^2}{\pi}}\left(\delta_{\mu\nu}-\frac{\left(k_
\mu\cdot k_\nu\right)}{k^2}\right)+\delta_{\mu\nu}\right\}.  \label{17}
\end{equation}

Collecting terms, one find that (\ref{12}) becomes
\begin{eqnarray}
&&4g^2\left\langle J_{\mu}(x)J_\nu(x^\prime)\right\rangle +
2\delta_{\mu\nu}\delta^2(x-x^\prime)\left\langle 1\right\rangle= \nonumber
\\
&=& \int  \frac{d^2k}{(2\pi)^2}\,G_{\mu\nu}^{-1}(k)e^{-ik(x-x^\prime)} . \label{18}
\end{eqnarray}

Thus, once (\ref{18}) is obtained, then the current algebra calculation is straightforward . Indeed,
from (\ref{18}) one see that the current-current commutator is
\begin{eqnarray}
&& \left< \left[ J_{\mu}(x_1),J_{\nu}(x^{\prime }_1)\right]\right>= \nonumber
\\
&=&\lim_{\tau\rightarrow 0^{+}}\frac{i}{2g^2}\int
\frac{d^2k}{2\pi} G_{\mu\nu}^{-1}(k)e^{-ik_1(x_1-x^{\prime}_ 1})\sin(k_0 \tau). \label{19}
\end{eqnarray}

In order to compute the RHS in (\ref{19}),  one note that in the limit $\tau \rightarrow 0^+$
becomes
\begin{eqnarray}
&& \lim_{\tau\rightarrow 0^+}\int dk_0\frac{k_0k_1}{k^{2}_0+ k^{2}_1}\sin(k_0\tau)= \nonumber
\\
&=&k_1\frac{\sqrt{\pi}}2\lim_{\tau\rightarrow 0^+}\tau\left( \frac{4k^{ 2}_1}{\tau^2}\right)^{\frac{1}{4}}
K_{\frac{1}{2}}(\sqrt{k^{ 2}_1\tau)}=\frac{\pi}{2} k_1
\end{eqnarray}
where the property
\begin{equation}
\lim_{z\rightarrow 0^+}K_{\frac{1}2}(z)=\frac{1}{2}\sqrt{\frac{2\pi}{z}}, \label{bb}
\end{equation}
has been used.

The calculation of the diagonal elements is straightforward and it gives
\begin{eqnarray}
&& \lim_{\tau\rightarrow 0^+}\int dk_0 \frac{k_0 k_0}{k^2}\sin(k_0\tau)= 0 \nonumber
\\
&& \lim_{\tau\rightarrow 0^+}\int dk_0\frac{k_1k_1}{k^2}\sin(k_0\tau)=0 \nonumber
\\
&&\lim_{\tau\rightarrow 0^+}\int dk_0\sin(k_0\tau)=0.  \label{20}
\end{eqnarray}

Therefore, the current algebra for the Thirring model becomes
\begin{equation}
\left\langle \left[J_{0}(x_1),J_{1}(x^{\prime }_1)\right]\right\rangle=\frac{c_T}{2\pi }
\delta^\prime(x_1-x^{\prime}_1), \label{201}
\end{equation}
where  the central charge $c_T$ is
\[
c_T = \frac{1}{1-\frac{(2g)^2}{2 \pi}}.
\]

One should note that the limit $g\rightarrow 0$ is smooth  and coincides with  the central
charge for free fermions.

However, there is a divergence for $c_T$  in  $g^2= \frac{\pi}{2}$. For this value of  $g^2$ our
calculation is no longer valid. Indeed, when $g^2= \frac{\pi}{2}$ one of the eigenvalues of
$G_{\mu \nu} (k)$ becomes negative and, therefore,  the calculation of the gaussian
integrals becomes meaningless.

Therefore,  the values $g^2 \geq \frac{\pi}{2}$ will  yields necessarily  inestabilities and the Thirring model is
only properly defined for  $c_T < 0$.

\section{Including Temperature and Density}

In this section we will explore the effects on the current algebra due to  finite temperature and
density.

By convenience we will work in the Minkowski space. In this case the Thirring lagrangian (\ref{9})
is changed by
\begin{equation}
{\cal L}_T -\mu\, \psi^{\dagger} \psi, \label{21}
\end{equation}
where $\mu$ is the chemical potential and the temperature effects are incorporated  in the
fermionic modified propagator, {\it v.i.z.}
\begin{equation}
S(k) \rightarrow S(k^{'})_{T=0} + S(k^{'})_{T\neq 0}, \label{22}
\end{equation}
where $k'_\nu=k_\nu+\mu ~u_\nu$ and $u_\nu$ being the four velocity with respect to the frame
where the heat bath is at rest.

In (\ref{22}) $S(k)_{T=0}$ is the standard Dirac propagator for massless fermions, but now
including the chemical potential contribution, {\it i.e.}
\begin{equation}
S(k)_{T=0} = \frac{k\s +\mu~ u\s }{(k+\mu~u)^2 +i~\epsilon}, \label{mud}
\end{equation}
while the temperature correction is
\begin{equation}
S(k)_{T\neq 0} = 2 i \pi (k\s +\mu~ u\s ) ~n_F(k\cdot u)~ \delta [(k+\mu~ u)^2 ],\label{tmud}
\end{equation}
where $n_F(k \cdot u)$ is the Fermi-Dirac distribution given by
\[
n_F(k) = \frac{1}{e^{\beta |k\cdot u|}+1}.
\]
 In the fluid  rest frame, the vector $u$ is $u_\mu =(1,0)$, and in such case, the Fermi-Dirac
 distribution depends only on $k_0$.

Contrarily to the zero temperature case, the effective action contains contributions to all orders
\cite{das} and, therefore, one could think that the finite temperature contributions could induce
corrections to  the central charge in the current algebra.

In order to explore this, let us consider the effective action up to the second order {\it i.e.}
\begin{equation}
S_{eff} = \int \frac{d^2k}{(2\pi)^2} A_\mu (k) \Pi_{\mu \nu} (k) A_\nu (-k), \label{23}
\end{equation}
where the polarization tensor is given by
\begin{eqnarray}
\Pi_{\mu \nu} (k) &=& - \frac{g^2}{2}\int \frac{d^2q}{(2\pi)^2} Tr \bigg[\gamma_\mu S(q) \gamma_\nu S(q-k)\bigg],
\label{24}
\\
&=& \Pi^{(0)}_{\mu \nu} (k) + \Pi^{(1)}_{\mu \nu} (k) + \Pi^{(2)}_{\mu \nu} (k) + \mbox{N.L.}, \label{244}
\end{eqnarray}
where $\mbox{N.L.}$ denote the non-local contributions that will be computed in the appendix but, they do not contribute to the central charge calculation.

The  equation (\ref{23}) yields three different contributions to the effective action. The first one,
formally,  is the
analogous to the $T=0$  case including now the chemical potential in the poles of $\Pi_{\mu\nu}^{(0)}$.
Naively, the presence of the chemical potential will produce a  non-trivial result.
However this is not true. Indeed, due to Lorentz invariance $\Pi_{\mu\nu}^{(0)}$ can be
decomposed according to
\[
\Pi_{\mu\nu}^{(0)} =  a~ \eta_{\mu \nu}\, k^2 + b~ k_\mu k_\nu.
\]

Additionally, the gauge invariance should imply that
\begin{equation}
k^\mu \Pi_{\mu\nu}^{(0)} =0, \label{nn}
\end{equation}
and then from (\ref{nn})  $a=-b$.

The explict form of  $a$ is obtained by computing  ${\Pi^{(0)}}_{\mu}^{\mu}$ {\it i.e.}
\begin{equation}
a= \Pi.  \label{kk}
\end{equation}

A straightforward calculation for ${\Pi^{(0)}}_{\mu}^{\mu}= \Pi$ yields
\[
a = \frac{g^2}{2 \pi},
\]
and, therefore, the chemical potential does not play any role in the first term of the effective
action.

This last fact is, of course, a two-dimensional  accident. Physically, this means that there is no compact matter in two-
dimensions.

The remaining diagrams could give temperature and density contributions.
The sum of the diagrams with one thermal insertion (fig. 1) are
\begin{figure}\label{suma}
    \epsffile{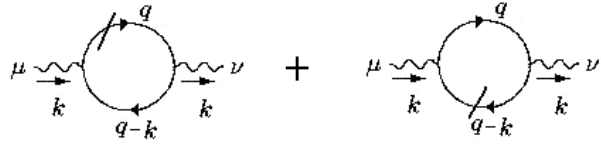}
    \centering
    {\caption{Diagrams with one thermal insertion}}
    \end{figure}

\begin{eqnarray}
&&\Pi^{(1)}_{\mu \nu} (k)= \Pi^{(12)}_{\mu \nu} (k) + \Pi^{(21)}_{\mu \nu} (k) \nonumber 
\\
&=&  \frac{g^2}{2}\int \frac{d^2q}{(2\pi)^2} Tr \bigg[\gamma_\mu~ S(q)_{T=0} ~\gamma_\nu ~S(q-k)_{T
  \neq 0}\bigg] \nonumber
\\
&+& \frac{g^2}{2}\int \frac{d^2q}{(2\pi)^2} Tr \bigg[\gamma_\mu~ S(q)_{T\neq 0} ~\gamma_\nu
~S(q-k)_{T= 0}\bigg],  \label{suma}
\end{eqnarray}

and explicitly each diagram is
\begin{eqnarray}
\pi i  \,g^2&&\int \frac{d^2q}{(2\pi)^2}\mbox{Tr}\left[\gamma_\mu \frac{(q\s +\mu u\s)}{(q~+\mu~u)^
2
+
i\epsilon}  \gamma_\nu (q\s-k\s+\mu~u\s)\right]\times\nonumber
\\
&\times& n_{\mbox{\scriptsize{f}}}(\beta| u\cdot (q-k)|)~\delta((q-k+\mu~u)^2), \label{di1}
\end{eqnarray}
and
\begin{eqnarray}
\pi i\,g^2\int\frac{d^2q}{(2\pi)^2}&{}&\mbox{Tr}\left[\gamma_\mu   (q\s+\mu~u\s )\gamma_\nu
\frac{(q\s-k\s+\mu~u\s)}{((q-k+\mu~u)^2+i\epsilon}
\right] \times \nonumber
\\
&\times&n_{\mbox{\scriptsize{f}}}(\beta| u\cdot q|) ~\delta((q+\mu~u)^2). \label{di2}
\end{eqnarray}

Note that both integrals are related. In fact, since they are convergent we can shift
variables. Indeed, the changes $q \rightarrow q-k+\mu~u$, for the first integral in (\ref{di1}) and $q \rightarrow q+\mu~u$ for the second one (\ref{di2}) imply that the previous integrals become 
\begin{eqnarray}
\pi i\,g^2\int&&\frac{d^2q}{(2\pi)^2}\mbox{Tr}\left[\gamma_\mu \frac{(q\s+k\s)}{(q+k)^2+i\epsilon}
\gamma_\nu~q\s\right]\times\nonumber
\\
&\times& n_{\mbox{\scriptsize{f}}}(\beta| u\cdot q-\mu|)~\delta(q^2), \label{di11}
\end{eqnarray}
and
\begin{eqnarray}
\pi i\,g^2\int\frac{d^2q}{(2\pi)^2}&{}&\mbox{Tr}\left[\gamma_\mu   q\s \gamma_\nu \frac{(q\s -k\s)}{((
q
-k)^
2+i\epsilon}
\right] \times \nonumber
\\
&\times&n_{\mbox{\scriptsize{f}}}(\beta| u\cdot q  -\mu|) ~\delta(q ^2). \label{di22}
\end{eqnarray}

Note now that,  the properties of gamma matrices  imply that
$$\mbox{Tr}[\gamma_\mu a\s \gamma_\nu b\s]=\mbox{Tr}[\gamma_\mu b\s \gamma_\nu a\s].$$

Therefore, the diagram corresponding to (\ref{di1}) is exactly the same corresponding to (\ref{di2})
by changing $k\rightarrow -k$. Equivalently, the diagram (\ref{di11})  is the same as (\ref{di22}) ($\Pi^{(12)}_{\mu \nu}(k)$) changing
$k\rightarrow -k$.

In order to compute one of these diagrams, firstly,  one can decompose the tensorial structure in (
\ref{di22}) as follows:
\begin{equation}
\Pi_{\mu\nu}^{(12)}(k)=\alpha^{(12)}\eta_{\mu\nu}+\beta^{(12)}\frac{k_\mu k_\nu}{k^2}+\gamma^{(
12)}\frac{k_{(\mu} u_{
\nu)}}{k\cdot u}, \label{albeto}
\end{equation}
where the notation $A_{(\alpha} B_{\beta)}$ stand for the symmetric product of $A$ and $B$.  One should note that any  2-rank symmetric tensor can be written as a linear combination of $\eta_{\mu \nu}$, $k_\mu k_\nu$ and $k_{(\mu} u_{\nu)}$, if $k_\mu$ and $u_\mu$ are independent. 

It is straightforward to prove that  the components $\alpha,\beta$ and $\gamma$ can be written in terms of Lorentz
invariants as
\begin{eqnarray}
\alpha^{(12)}&=&\Pi_{\mu\nu}^{(12)}\eta^{\mu\nu}-\Pi_{\mu\nu}^{(12)}\frac{k^\mu k^\nu}{k^2},
\nonumber
\\
\beta^{(12)}&=&3\Pi_{\mu\nu}^{(12)}\eta^{\mu\nu} +\left(\frac{1}{(k\cdot u)^2-k^2}\right)\bigg[(k\cdot
u)\Pi_{\mu\nu}^{(12)}
k^{(\mu}u^{\nu)}+\nonumber
\\
&& \Pi_{\mu\nu}^{(12)}k^\mu k^\nu\bigg],\nonumber
\\
\gamma^{(12)}&=&\frac{1}{2}\left(\frac{(k\cdot u)^2}{(k\cdot u)^2-k^2}\right)\left[2\Pi_{\mu\nu}^{(12)}
\frac{k^\mu
k^\nu}{k^2}-
\Pi_{\mu\nu}^{(12)} \frac{k^{(\mu}u^{\nu)}}{k\cdot u}\right]
\end{eqnarray}

Therefore, the only relevant integrals that must be performed are the corresponding to
$\Pi_{\mu\nu}^{(12)} \eta^{\mu\nu}$, $\Pi_{\mu\nu}^{(12)} k^\mu k^\nu$ and
$\Pi_{\mu\nu}^{(12)} k^{(\mu}u^{\nu)}$.

Straightforwardly one can prove that $\Pi_{\mu\nu}^{(12)} \eta^{\mu\nu}=0$. The product
$ \Pi_{\mu\nu}^{(12)}k^\mu k^ \nu$ is
\begin{eqnarray}
&&\Pi_{\mu\nu}^{(12)}k^\mu k^\nu =\nonumber
\\
&&2\pi ig^2\,\int\frac{d^2q}{(2\pi)^2}[2(q\cdot k)((q-k)\cdot k)-k^2(q-k)\cdot q]\times\nonumber
\\
&&\times \frac{n_{\mbox{\scriptsize{f}}}(\beta (|u\cdot q-\mu|))\delta(q^2)}{(q-k)^2+i\epsilon},
\nonumber
\\
&=&2\pi i\,g^2 \,\int\frac{d^2q}{(2\pi)^2}(q\cdot k)\frac{2(q\cdot k)-k^2}{(q-k)^2+i\epsilon} n_{\mbox{
\scriptsize{f}}}(\beta |u\cdot q-\mu|)\delta(q^2),\nonumber
\\
&=&- 2\pi ig^2\, \int\frac{d^2q}{(2\pi)^2}(k\cdot q)n_{\mbox{\scriptsize{f}}}(\beta| u\cdot q-\mu|)\delta
(q^2).
\end{eqnarray}

This trace can be recast in a more transparent fashion by rescaling the integration variable
$\beta q \rightarrow q$.

This trace becomes
\begin{equation}
\Pi_{\mu\nu}^{(12)}k^\mu k^\nu=-2\pi ig^2 \frac{k^\alpha v_\alpha}{\beta} ,
\end{equation}
where
$$v_\alpha=\int\frac{d^2q}{(2\pi)^2}q_\alpha~n_{\mbox{\scriptsize{f}}}(| u\cdot q-\beta\mu|)\delta(q^
2).$$

Finally, let us point out that the vector $v_\alpha$ has components only along the direction
defined by $u_\mu$. That
means that
\begin{equation}
\Pi_{\mu\nu}^{(12)}k^\mu k^\nu=-4\pi i\frac{k\cdot u}{\beta} \int\frac{d^2q}{(2\pi)^2}(q\cdot u)~n_{
\mbox{\scriptsize{f}}}(| u\cdot q-\beta\mu|)\delta(q^2).
\end{equation}

In order to compute the coefficient in front of  $k^{(\mu}u^{\nu)}$, one proceed as follows: firstly, one compute
\begin{eqnarray}
\Pi_{\mu\nu}^{(12)}k^{(\mu} u^{\nu)}&=&4\pi ig^2
\int\frac{d^2q}{(2\pi)^2}\frac{n_{\mbox{\scriptsize{f}}}(
\beta |u\cdot q-\mu|)\delta(q^2)}{(q-k)^2+i\epsilon}\nonumber
\\
&&[(q\cdot k)((q-k)\cdot u)+(q\cdot u)((q-k)\cdot k)-\nonumber
\\
&&(k\cdot u)((q-k)\cdot q)],\nonumber
\\
&=&4\pi ig^2 \int\frac{d^2q}{(2\pi)^2}(q\cdot u)\frac{2(q\cdot k)-k^2}{(q-k)^2+i\epsilon}\times
\nonumber
\\
&&n_{\mbox{\scriptsize{f}}}(\beta |u\cdot q -\mu|)\delta(q^2),\nonumber
\\
&=&- 4\pi ig^2 \int\frac{d^2q}{(2\pi)^2}(q\cdot u)n_{\mbox{\scriptsize{f}}}(\beta| u\cdot
q-\mu|)\delta(q^2), \nonumber
\\
\end{eqnarray}
then, this last  integral can be recast by introducing the same rescaling of q as
before.
\begin{equation}
\Pi_{\mu\nu}^{(12)}k^{(\mu} u^{\nu)}=\frac{-4\pi ig^2}{\beta} \int\frac{d^2q}{(2\pi)^2}(q\cdot u)~n_{
\mbox
{\scriptsize{f}}}(| u\cdot q-\beta\mu|)\delta(q^2).
\end{equation}

The coefficients $\alpha^{(12)},\cdots$ can be evaluated directly, indeed. One find
\begin{eqnarray}
\alpha^{(12)}&=&2\pi ig^2\frac{k\cdot u}{\beta k^2}\lambda(\beta\mu,u),
\\
\beta^{(12)}&=&0,
\\
\gamma^{(12)}&=&-2\pi ig^2\frac{k\cdot u}{\beta k^2}\lambda(\beta\mu,u).
\end{eqnarray}
where
\[
\lambda(\beta\mu,u)=\int\frac{d^2q}{(2\pi)^2}(q\cdot u)~n_{\mbox{\scriptsize{f}}}(| u\cdot q-\beta\mu
|)\delta(q^2).
\]

Therefore
\begin{equation}
\Pi_{\mu\nu}^{(12)}(k)=2\pi ig^2\frac{k\cdot u}{\beta k^2}\lambda(\beta\mu,u)\left(\eta_{\mu\nu}-
\frac{k_
{(\mu}u_\nu) }{k\cdot u}\right) \nonumber
\end{equation}

From this last result we see that this diagram is odd respect to the change $k\rightarrow -k$.
Therefore, the sum of diagrams (\ref{di1}) and (\ref{di2}) is zero
\begin{eqnarray}
\Pi^{(12)}_{\mu\nu}(k)+\Pi^{(21)}_{\mu\nu}(k)&=&\Pi^{(12)}_{\mu\nu}(k)+\Pi^{(12)}_{\mu\nu}(-k)
\nonumber
\\
&=&\Pi^{(12)}_{\mu\nu}(k)-\Pi^{(12)}_{\mu\nu}(k)\nonumber
\\
&=&0.
\end{eqnarray}

Finally, we must compute the diagram with two thermal s, (fig. 2)
\begin{figure}[t]\label{suma}
 \centering 
    \epsffile[-2.2cm -.1cm 1cm 2.5cm ]{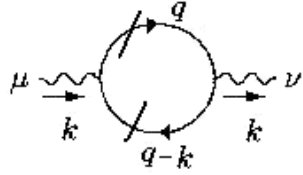}
    \caption{Diagram with two thermal insertions}
    \end{figure}
    given by the expression
    \begin{eqnarray}
    \Pi_{\mu\nu}^{(2)}(k)&=&2\pi ig^2\int\frac{d^2q}{(2\pi)^2}\times \nonumber 
    \\
    &\times& n_{\mbox{\scriptsize{f}}}[\beta |u\cdot(k-
    q)- \mu|]
    ~n_{\mbox{\scriptsize{
    f}}}[\beta |u\cdot q -\mu|]\delta(q^2)\nonumber
    \\
    &&\hspace{-2cm}\delta((k-q)^2)[q_\mu (q -k )_\nu+ q_\nu (q -k )_\mu-\eta_{\mu\nu}(q-k)\cdot q].
\end{eqnarray}
 it turns out that
$$ \Pi_{\mu\nu}^{(2)}(k)=0$$

Once these diagramas are computed,  then the effective action becomes like (\ref{23}), but with
poles in $k_0$ shifted to $k_0-\mu$. 

In the case of gauge theories like QCD, the cancellation of these thermal diagrams is a general
and mandatory  property if the gauge invariance is preserved.

The calculation of the current algebra for the Thirring model is straightforward and, as a
consequence,  there are no modifications due to finite temperature and density effects in two
dimensions.  In the next section we will come back to this conclusion when we go beyond two dimensions

\section{Finite Temperature and density beyond  two-dimensions}

In this section we argue how to extend our results to the four dimensional case. One start
considering the lagrangean
\begin{equation}
{\cal L} = \bar\psi\left(i\partial\hspace{-2.4mm}\slash\hspace{1.4mm}-g{\cal B}\hspace{-2.4mm}
\slash
\hspace{1.4mm} -m \right)\psi - \frac{1}{4} F^2, \label{300}
\end{equation}
where the effective gauge field  ${\cal B}$ is
\begin{equation}
{\cal B}_\mu = A_\mu + \frac{1}{g} v_\mu, \label{3000}
\end{equation}
 $v_\mu$ the velocity of the fluid and $m$ an ultraviolet regulator.

The effective action for this case is
\begin{equation}
S_{eff} = \int \frac{d^4k}{(2\pi)^2} {\cal B}_\mu (k) \Pi_{\mu \nu} (k) {\cal B}_\nu (-k), \label{301}
\end{equation}
where the polarization tensor
\begin{equation}
\Pi_{\mu \nu} = \Pi^{(0)}_{\mu \nu}+\Pi^{(1)}_{\mu \nu}+\Pi^{(2)}_{\mu \nu}, \label{302}
\end{equation}
is now quadratically divergent and and it needs to be renormalized. The contributions in (\ref{302})
include the standard self-energy gauge field diagram plus the contributions with one and two
thermal insertions.

In analogy with our previous calculation, one could conjecture that these thermal contributions
should vanishes due to gauge invariance of the effective action (\ref{301}). However, (\ref{albeto}) does not remain valid in the 4-dimensional 
because we have to introduce a new contribution proportional to $u_\mu u^\mu$ \cite{dit}. 

However, the contribution $\Pi^{(0)}_{\mu \nu}$  could contain  chemical potential
contributions.
Actually,  this effective action written in terms of the original fields becomes
\begin{eqnarray}
&&S_ {eff} = g^2 \int d^4~k A_\mu (k) \Pi^{(0) \mu \nu} A_\nu (-k) \nonumber
\\
&+& 2 g\int d^4k ~A_\mu (k) \Pi^{(0) \mu \nu} ~v_\nu + g \int d^4k~ v_\mu (k) \Pi^{(0)\mu \nu} v_
\nu.  \nonumber
\\
\label{303}
\end{eqnarray}

The renormalized polarization tensor  is
\begin{equation}
\Pi^{(0)}_{\mu \nu} = \left(q^2 ~\eta_{\mu \nu} - q_\mu q_\nu\right) \Pi^{(0)} (q^2), \label{304}
\end{equation}
where $\Pi^{(0)}(q^2)$ is a finite contribution given by
\begin{eqnarray}
\Pi^{(0)}(q^2) &=& -\frac{g^2}{2 \pi^2} \int_0^1 dx ~x(1-x) \ln \left( \frac{m^2}{m^2 - x(1-x)q^2}\right)
\nonumber
\\
&=&\frac{1}{q^3}\log \,m^2\,\left( -q + \frac{4\,m^2\,
         \arctan (\frac{q}{{\sqrt{4\,m^2 - q^2}}})}{{\sqrt{4\,m^2 - q^2}}} \right) \nonumber
         \\
   \label{fini}
\end{eqnarray}
where $m^2$ is a cut-off which is necessary in order to have a well defined integral.

The diagram implies, as usual, charge renormalization where the
effects induced by density and temperature appears in the $q_0$ component if the substitution
$q_0 -\mu$  is understood.

Thus, one can conclude that the effective action can, of course, depend of the temperature and
density and that the gauge invariance is a preserved symmetry.

In the case of the current algebra calculation the situation is different. Indeed,  corrections to the
central charge come from the term
$k_\mu k_\nu$ and as this term is multiplied by $\Pi (q^2)$, then the central charge would be
modified by density and finite temperature.

We will analyses this problem elsewhere.
\acknowledgments
We would like to thank  Prof. A. Das for discussions. This work has been partially supported by
the grants 1010596, 1010976 from Fondecyt-Chile and Mecesup-USA 109 and Spanish Ministerio de Educaci\'on y Cultura grant- FPU AP07566588 and 
O.N.C.E. for support.  

\appendix 
\section{One-loop Non-local Corrections to the effective action}

At the one-loop order, the polarization tensor in components can be written as follows 
({\it v.i.z.} (\ref{24})):
\begin{eqnarray}
\Pi^{00}_\beta(k)&=&\Pi^{11}_\beta(k)=2\int\frac{d^2q}{(2\pi)^2}(q^0(k+q)^0+q^1(q+k)^1) 
\left(\frac{1}{k^2+i\epsilon}+(2\pi i)n_F(\vert k\vert)\delta(k^2)\right)\times\nonumber\\
&\times&\left(\frac{1}{(q+k)^2+i\epsilon}+(2\pi i)n_F(\vert k+q\vert)\delta((q+k)^2)\right)
\label{primer2}
\end{eqnarray}
and,
\begin{eqnarray}
\Pi^{01}_\beta(k)&=&\Pi^{10}_\beta(k)=2\int \frac{d^2q}{(2\pi)^2}\,(q^0(k+q)^1-q^1(q+k)^0) \left(\frac{1}{k^2+i\epsilon}+(2\pi i)n_F(\vert k\vert)\delta(k^2)\right)\times\nonumber\\
&\times& \left(\frac{1}{(q+k)^2+i\epsilon}+(2\pi i)n_F(\vert k+q\vert)\delta((q+k)^2)\right). 
\label{primer3}
\end{eqnarray}

Our next step is to compute the diagonal thermal corrections of $\Pi^{\mu \nu}$. The first correction comes from  
\begin{equation}
4\pi i\int\frac{d^2q}{(2\pi)^2}\frac{q^0(q+k)^0+q^1(q+k)^1}{(k^0)^2-(k^1)^2+i\epsilon}\delta((q+k)^2)n_F(\vert k^0+q^0\vert). 
\label{segun}
\end{equation}

However, before to computing (\ref{segun}), it is convenient to use the identity 
\begin{equation}
\delta((q+k)^2)=\frac{1}{2\vert k^1+q^1\vert}\left(\delta(q^0+k^0-(q^1+k^1))+
\delta(q^0+k^0+(q^1+k^1)\right). 
\end{equation}

Therefore, (\ref{segun}) becomes 
\begin{eqnarray}
2\pi i\sum_\pm\int\frac{d^2q}{(2\pi)^2}\frac{2q^1+(k^1\mp k^0)}{(2q^1+ (k^1\mp k^0))(k^1\mp k^0)+i\epsilon}\times&\nonumber\\
&\hskip -5cm\times\epsilon(k^0+q^0)\delta(q^0-k^0\pm(q^1+k^1))n_F(\vert q^0+k^0\vert). 
 \label{segun1}
\end{eqnarray}

We will make use the Plemelj's decomposition to compute these integrals:
$$\frac{\alpha}{\alpha (x-x_0)+i\epsilon}={\cal P} \left(\frac{1}{x-x_0}\right)-
i\epsilon(\alpha)\pi\delta(x-x_0)$$
where ${\cal P}$ mean the principal value and the $\epsilon$-prescrition is assumed.  

Using this identity, (\ref{segun1})  becomes 
\begin{equation}
-2\pi i\left(\frac{2k^1}{k^2}+\pi i\left[\delta(k^0-k^1)+\delta(k^0+k^1)\right]\right) \int\frac{dq^1}{(2\pi)^2}
\epsilon(q^1)\epsilon(q^1+k^1) n_F(\vert k^1+q^1\vert).
\label{insercion1}
\end{equation}

The next correction to the diagonal terms is calculated in a similar way. Indeed, if we use  
\[
\delta(q^2)=\frac{1}{2\vert q^1\vert}\left(\delta(q^0+q^1)+\delta(q^0-q^1)\right), 
\]

One find that 

\begin{eqnarray}
-2\pi i\left(-\frac{2k^1}{k^2}+\pi i\left[\delta(k^0-k^1)+\delta(k^0+k^1)\right]\right)\times & \nonumber\\
 &\hskip -3cm\times\int\frac{dq^1}{(2\pi)^2}
\epsilon(q^1)\epsilon(q^1+k^1) n_F(\vert q^1\vert).
\label{insercion2}
\end{eqnarray}

Finally, the last contribution including two thermal insertions is 
 \begin{eqnarray}
2(2\pi i)^2\int\frac{d^2q}{(2\pi)^2}\left(q^0(q^0+k^0)+q^1(q^1+k^1)\right)\times&\nonumber\\
&\hskip -8cm\times\frac{1}{4\vert (q^1+k^1)k^1\vert}\left(\delta(q^0+k^0-(q^1+k^1))+\delta(q^0+k^0+(q^1+k^1))\right)\times\nonumber\\
&\hskip -6cm\times\left(\delta(q^0-q^1)+\delta(q^0+q^1)\right) n_F(\vert q^0\vert)n_F(\vert q^0+k^0\vert). 
 \end{eqnarray}
 
After collecting terms, the non-local corresctios to the diagonal terms is given by 
\begin{eqnarray}
\Pi^{00}_\beta(k) &=& 2\pi^2\left(\delta(k^0+k^1)+\delta(k^0-k^1)\right) 
\\
&\times&  
\int\frac{d^2q}{(2\pi)^2}\epsilon(q^1)\epsilon(q^1+k^1) \biggl\{n_F(\vert q^1\vert)+n_F(\vert q^1+ k^1  \vert) - 2n_F(\vert q^1\vert)n_F(\vert q^1+k^1\vert)\biggr\}
\label{correccion}
\end{eqnarray}
in full agreement with Das and Da Silva \cite{das}.  Of course, these contributions do not modify the central charge. 

In the same way, one verify that 
\begin{eqnarray}
\Pi^{01}_\beta(k)&=&2\pi^2\left(\delta(k^0+k^1)-\delta(k^0-k^1)\right) \int\frac{d^2q}{(2\pi)^2}\epsilon(q^1)\epsilon(q^1+k^1)\times \nonumber 
\\
&\times&\left\{n_F(\vert q^1\vert)+n_F(\vert q^1+k^1\vert)-2n_F(\vert q^1\vert)n_F(\vert q^1+k^1\vert)\right\}.
\label{correccion1}
\end{eqnarray}

All these results are proportional to the integral 
\begin{eqnarray}
I_2(k^1;\beta)=\int\frac{dq^1}{(2\pi)^2}\left[\epsilon(q^1)\epsilon(k^1+q^1)\right.\times&\nonumber\\
&\hskip -7cm\times\left.\left\{n_F(\vert q^1\vert)+n_F(\vert k^1+q^1\vert)-2n_F(\vert q^1\vert)n_F(\vert
k^1+q^1\vert)\right\}\right]. \label{vamp}
\end{eqnarray}

Although  (\ref{vamp}) is a cumbersome expression, the limit $\beta \rightarrow \infty$ can be easily computed. Indeed, since $I_2\rightarrow 0$ when $\beta\rightarrow\infty$, then in this limit 
\begin{equation}
I_2=\frac{1}{(2\pi)^2}\int_{-\infty}^\infty dq^1\,\epsilon(q^1)\epsilon(k^1+q^1)
\frac{\cosh\left(\beta\frac{\vert q^1\vert-\vert k^1+q^1\vert}{2}\right)}{\cosh\left(
\beta\frac{\vert q^1\vert+\vert k^1+q^1\vert}{2}\right)+\cosh\left(\beta\frac{\vert q^1\vert-\vert k^1+q^1\vert}{2}\right)}. 
\label{ooo}
\end{equation}

If we assume that $p^1>0$ and one consider three regions on the real axis, namely,  ${\cal R}_1=(-\infty, -p^1)$, ${\cal R}_2=(-p^1,0)$ and  ${\cal R}_3=(0,\infty)$. Then in ${\cal R}_1$ and ${\cal R}_3$ the product $\epsilon(q^1)\epsilon(k^1+q^1)$ is positive and in ${\cal R}_2$ is negative. 

Thus in these regions we have: 
\begin{eqnarray}
\vert q^1\vert=-q^1,&\vert k^1+q^1\vert=-(k^1+q^1)&{\rm en}\,{\cal R}_1\nonumber\\
\vert q^1\vert=-q^1,&\vert k^1+q^1\vert=(k^1+q^1)&{\rm en}\,{\cal R}_2\nonumber\\
\vert q^1\vert=q^1,&\vert k^1+q^1\vert=(k^1+q^1)&{\rm en}\,{\cal R}_3\nonumber
\end{eqnarray}
and, therefore, 
\[
I_2=
\frac{1}{(2\pi)^2}\left(\int_{{\cal R}_1\cup{\cal R}_3}dq^1\,\frac{\cosh(\beta k^1/2)}{\cosh(\beta k^1/2)+\cosh(\beta(k^1+q^1/2))}
 -\int_{{\cal R}_2}dq^1\,\frac{\cosh(\beta(k^1+q^1/2))}{\cosh(\beta(k^q+ q^1/2))+\cosh(\beta k^1/2)}\right).
\label{oooo}
\]

However, in the last integral one use the identity  
\begin{equation}
\frac{\cosh(\beta(k^1+q^1/2))}{\cosh(\beta(k^1+q^1/2))+\cosh(\beta k^1/2)}=1-
\frac{\cosh(\beta k^1/2)}{\cosh(\beta k^1/2)+\cosh(\beta(k^1+q^1/2))}, 
\end{equation}
and, therefore, one obtain that 

\begin{equation}
I_2=-\frac{1}{(2\pi)^2}\left(k^1-\frac{1}{\beta}\int_{-\infty}^\infty dx\,\frac{\cosh(\beta k^1/2)}{\cosh(\beta k^1/2)+\cosh(x)}\right)
\label{aa}
\end{equation}
where in the last integral the change of variables $x=\beta(k^1+q^1/2)$ was performed.

Finally, $I_2$ is 
\begin{equation}
I_2(k^1;\beta)=-\frac{1}{(2\pi)^2}k^1(1-{\rm cotgh}(\beta k^1/2))=-e^2\frac{k^1}{e^{\beta`k^1}-1}
\label{final}, 
\end{equation}
where at low temperatures its behavior is given by 
$$I_2(k^1,\beta)\sim \frac{1}{8\pi^2}k^1e^{-\beta k^1}$$
in full agreement with the expected result \cite{das}.


\begin{references}
 \bibitem{pes} See {\it e.g.} M. E. Peskin and D. V. Schroeder, {\it An Introduction to Quantum
 Field Theory}, Addison-Wesley 1997.
 \bibitem{ka}  J. Kapusta {\it Finite-temperature field theory}, Cambridge Monographs on
 Mathematical Physics, Cambridge University Press, 1989. M. Le Bellac {\it Thermal Field Theory},
 Cambridge Monographs on Mathematical Physics, Cambridge University Press, (1996).
  \bibitem{lo} M. Loewe and  J. C. Rojas,  {\it Phys. Rev.} {\bf D46} 2689 (1992).
  \bibitem{we} H. A. Weldon, {\it Phys. Rev.} {\bf D26}, 1394 (1982);   A. Actor, {\it Phys. Rev.} {\bf D 27}, 2548 (1983).
  \bibitem{fp} F. Pauquay, hep-ph-/0212080; D. Boyanovsky, H. J. Vega, H. Holman,  and M. Dimionato, {\it Phys. Rev.} {\bf D60}, 065003 (1999).
 \bibitem{da1} A. Das, {\it Topics in finite temperature field theory.}. Published in {\it Quantum Field
 Theory - A Twentieth Century Profile}, Indian National Science Academy.  {\it Quantum field
 theory* 383-411}, Mitra, A.N. (ed.){\it hep-ph/0004125}.
 \bibitem{re}  For a general review see also,A. Das, {Finite Temperature Field Theory}, World Scientific (1997). 
 \bibitem{joh} The original calculation of the current algebra are considered in: K. Johnson, {\it Nuovo Cim.} {\bf 20}, 773 (1961); C. Sommerfield, {\it Ann. Phys.} {\bf 26}, 1 (1964); C. R. Hagen, 
 {\it Nuovo Cim.} {\bf 51B}, 169 (1967). But these authors consider the current  algebra for the case $T=0$ and $\mu =0$.
 \bibitem{elcio} This calculation is performed using bosonization in {\it i.e.} E. Abdalla, M. C. Abdalla and K. Rothe, {\it 2  Dimensional Quantum Field Theory}, World Scientific (1991); the most important references are collected in, M. Stone, {\it Bosonization}, World Scientific 1994.
 \bibitem{stone}  The most important references are collected in, M. Stone, {\it Bosonization},
 World Scientific 1994.
 \bibitem{fuckiw} See {\it e.g.} R. Jackiw in {\it Current Algebra and Anomalies}, World Scientific, 1985.   
 \bibitem{das}  A. Das and A. J. da Silva, {\it  Phys.Rev.} {\bf D59}, 105011 (1999) and references
 therein.
 \bibitem{dit} W. Dittrich, H. Gies, {\it Probing the Quantum Vacuum: Perturbative Effective Action Approach in Quantum Electrodynamics and Its Applications} , pp 115. Springer-Verlag  (2000).






   \end{references}
\end{document}